**SGUQ: Staged Graph Convolution Neural Network for Alzheimer's Disease Diagnosis using Multi-Omics Data**

Liang Tao[1], Yixin Xie[2], Jeffrey D Deng[3], Hui Shen[4], Hong-Wen Deng[4], Weihua Zhou[5,6], Chen Zhao[1]*

1. Department of Computer Science, Kennesaw State University, Marietta, GA 30060

2. Department of Information Technology, Kennesaw State University, Marietta, GA, 30060

3. Geisel School of Medicine at Dartmouth College, Hamover, NH 03755

4. Division of Biomedical Informatics and Genomics, Tulane Center of Biomedical Informatics and Genomics, Deming Department of Medicine, Tulane University, New Orleans, LA 70112

5. Department of Applied Computing, Michigan Technological University, Houghton, MI, 49931

6. Center for Biocomputing and Digital Health, Institute of Computing and Cybersystems, and Health Research Institute, Michigan Technological University, Houghton, MI 49931

* Corresponding author:

Chen Zhao, PhD

Department of Computer Science, Kennesaw State University,

680 Arntson Dr, Atrium BLDG, Marietta, GA 30060Marietta, GA 30060

Email: czhao4@kennesaw.edu

**Abstract**

Alzheimer's disease (AD) is a chronic neurodegenerative disorder and the leading cause of dementia, significantly impacting cost, mortality, and burden worldwide. The advent of high-throughput omics technologies, such as genomics, transcriptomics, proteomics, and epigenomics, has revolutionized the molecular understanding of AD. Conventional AI approaches typically require the completion of all omics data at the outset to achieve optimal AD diagnosis, which are inefficient and may be unnecessary. To reduce the clinical cost and improve the accuracy of AD diagnosis using multi-omics data, we propose a novel staged graph convolutional network with uncertainty quantification (SGUQ). SGUQ begins with mRNA and progressively incorporates DNA methylation and miRNA data only when necessary, reducing overall costs and exposure to harmful tests. Experimental results indicate that 46.23% of the samples can be reliably predicted using only single-modal omics data (mRNA), while an additional 16.04% of the samples can achieve reliable predictions when combining two omics data types (mRNA + DNA methylation). In addition, the proposed staged SGUQ achieved an accuracy of 0.858 on ROSMAP dataset, which outperformed existing methods significantly. The proposed SGUQ can not only be applied to AD diagnosis using multi-omics data, but also has the potential for clinical decision making using multi-viewed data. Our implementation is publicly available at https://github.com/chenzhao2023/multiomicsuncertainty.

**Keywords**: Alzheimer's Disease, Staged decision making, Multi-omics information fusion, Uncertainty quantification

# 1. Introduction

Alzheimer's disease (AD) is a chronic neurodegenerative disorder and the leading cause of dementia, significantly impacting cost, mortality, and burden in the United States [1,2]. Despite its prevalence, the exact etiology of AD remains unclear. Identifying risk factors for AD is crucial for understanding its pathogenesis and improving diagnosis and treatment. Additionally, genomic data, can provide valuable insights into genetic predispositions, such as APOE ε4, CLU and CR1, which increases AD risk [3,4].

The advent of high-throughput omics technologies-such as genomics, transcriptomics, proteomics, and epigenomics-has revolutionized our understanding of biological systems at the molecular level for AD diagnosis [5,6]. However, each omics technique provides only a fragmented view of complex biological processes. Integrating heterogeneous omics data has been shown to yield a comprehensive understanding of diseases and phenotypes [7]. Multi-omics integration has demonstrated significant improvements in disease prediction accuracy and biomarker discovery [8]. Studies have shown that combining data from multiple omics layers enhances patient clinical outcome predictions compared to using a single data type [9,10].

Despite these advancements, collecting comprehensive multi-omics data is often accompanied by substantial costs and technical challenges. High-throughput experiments are expensive and resource-intensive, limiting their feasibility in large-scale clinical applications or resource-constrained settings [11]. Traditional approaches require completing all diagnostic tests upfront, which increases costs and inefficiencies, while failing to incorporate a staged, cost-effective method for prioritizing less effective or expensive diagnostic tools [12]. In contrast to the existing methods, we propose to develop an AI-driven, staged approach for AD diagnosis that prioritizes less expensive and most effective tests initially, escalating to more sophisticated tools only when necessary to obtain a conclusive result.

Early supervised data integration methods, such as feature concatenation and ensemble strategies [9], often failed to consider correlations among different omics types and could be biased toward certain data modalities. Recent methods have focused on exploiting interactions across omics types using advanced machine learning techniques, including neural networks [7,8,13], graph-based approaches [12], pathway-based integration [14] and variational autoencoders [14]. For example, some studies have utilized self-attention mechanisms and contrastive learning to dynamically identify informative features and maximize mutual information between different omics types, enhancing classification performance even when data is incomplete [11]. However, many existing methods still face challenges such as information loss due to data heterogeneity and computational difficulties arising from high data dimensionality [15]. Therefore, developing cost-effective methods capable of maximizing predictive accuracy while minimizing the required amount of omics data is essential.

In this paper, we propose developing a novel staged graph convolutional network with uncertainty quantification (SGUQ) for AD risk prediction using multi-omics data integration. By considering the graph connectivity, we ensure that critical relationships and interdependencies between various omics data are maintained, allowing for a more holistic and accurate modeling of the disease pathology. Our SGUQ begins with the most informative omics data and only incorporates necessary tests for a definitive assessment. An uncertainty quantification model is proposed alongside the AD diagnosis model to determine whether additional screening is necessary. Experimental results indicate that an accuracy of 0.858 was achieved for AD diagnosis, while 46.23% patients require only mRNA omics data.

# 2. Methodology

In this section, we introduce the proposed SGUQ for AD diagnosis, which is framed as a binary classification task. The overview of SGUQ is shown in Figure 1.

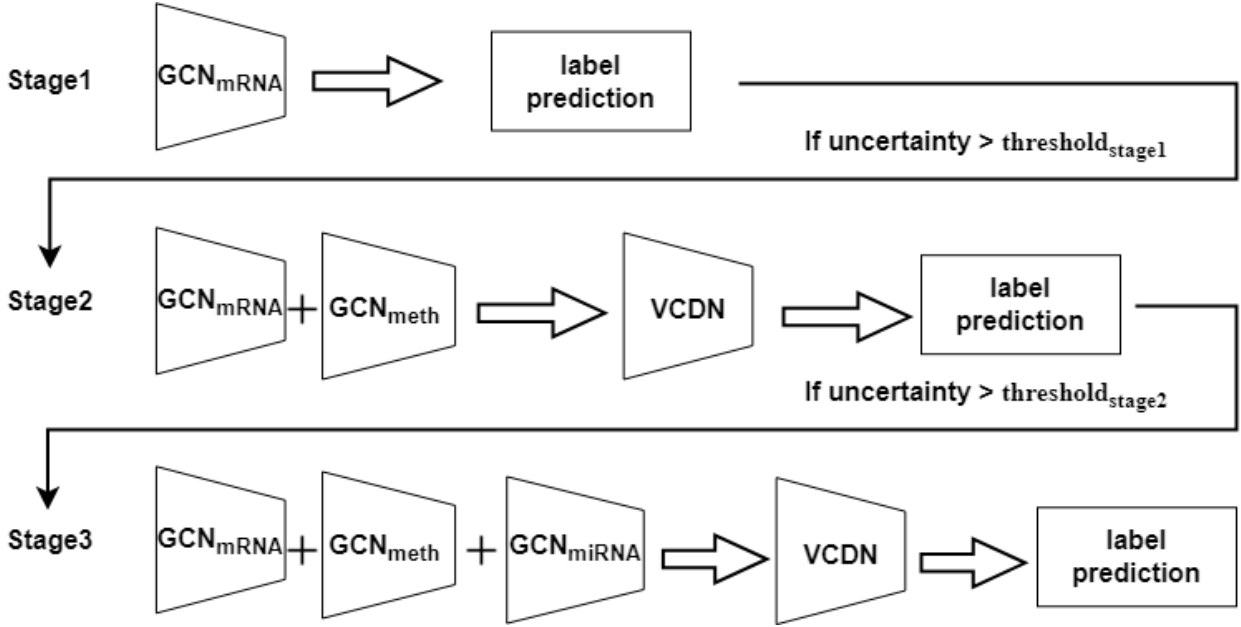

**Figure 1**. Overview of SGUQ. Our staged method employs the most informative omics data to perform AD diagnosis while evaluating the uncertainty. In this example, at stage 1, only mRNA data are applied for prediction. At the following stages, the uncertainty is evaluated to decide whether more omics data (DNA methylation, miRNA) need to be introduced to make a definitive decision. Once the uncertainty is smaller than the threshold, the prediction is terminated.

**2.1 AD classification using single omics data**

In multi-omics data analysis, identifying the most effective single omics modality is a critical first step before considering the integration of multiple data types. In our SGUQ, we build specialized graph convolutional networks (GCNs) [16] for each omics data type to perform classification tasks. Each GCN is trained on a single omics modality and aims to capture the intrinsic structure of the data, thereby improving classification performance. In a supervised learning context, the benefit of using GCN is that it not only captures local internal information within the same category due to multi-modal phenomena, but also finds more discriminative features by considering information between similar patients. By evaluating each omics modality individually using the accuracy score, we establish a baseline performance that reflects the predictive power of each data type on its own.

In a single-view model, each sample is regarded as a node in the sample similarity network, and the edges between nodes represent the similarity between samples. Suppose that the omics feature for one patient is defined as $X_i^{(k)} \in \mathbb{R}^{d_k}$, where $i$ is the index of the patient, $k$ is the index of the omics and $d_k$ represents the dimension of the features for $k$-th omics data. We aim to use GCN to perform feature representation learning while considering the graph connectivity, and use the extracted features to perform AD diagnosis. Under this scenario, each GCN requires the following two inputs: a feature matrix $X^{(k)} \in \mathbb{R}^{n \times d_k}$, where $X^{(k)}$ represents the omics data for $n$ samples for the $k$-th omics data; and an adjacency matrix $A \in \mathbb{R}^{n \times n}$. The adjacency matrix is constructed based on the cosine similarity between patients. For each pair of samples $X_i^{(k)}$ and $X_j^{(k)}$, we calculate the cosine similarity of their feature vectors as the graph connectivity, as shown in Eq. 1.

$$s\left(X_i^{(k)}, X_j^{(k)}\right) = \frac{X_i^{(k)} \cdot X_j^{(k)}}{||X_i^{(k)}||_2 \cdot ||X_j^{(k)}||_2} \quad (1)$$

A threshold $\epsilon$ is set so that connections are established between patient pairs only if their similarity is greater than or equal to $\epsilon$. The selection of the threshold $\epsilon$ is based on the parameter $K^{(k)}$ for the $k$-th omics, which is the average number of edges retained per node (including self-connections), as defined in Eq. 2. By adjusting the value of $K$ (e.g., 2, 5, 10), we can control the sparsity of the graph:

$$K^{(k)} = \frac{\sum_{i,j} \mathbb{I}(s(X_i^{(k)}, X_j^{(k)}) \geq \epsilon)}{n} \quad (2)$$

where $\mathbb{I}(\cdot)$ is an indicator function. Note that if $K^{(k)} = 1$, no edge is enrolled, then the adjacency matrix will only include self-connections, and GCN will degenerate to a fully connect network. As a result, the element of the adjacency matrix is measure in Eq. 3.

$$A_{ij}^{(k)} = \begin{cases} s(X_i^{(k)}, X_j^{(k)}), & if\ i \neq j\ and\ s(X_i^{(k)}, X_j^{(k)}) \geq \epsilon \\ 0, & Otherwise \end{cases} \quad (3)$$

In SQUQ, feature extraction is performed by stacking multiple GCN layers. The propagation rules of each layer are defined in Eq. 4.

$$H^{(l+1)} = \sigma(\tilde{A} H^{(l)} W^{(l)}) \quad (4)$$

where $H^{(l)}$ is the input feature matrix of the $l$-th layer. $l \in [0, \cdots, L]$ represents the index of GCN layer, and $L$ is the total number of layers employed in GCN. $W^{(l)}$ is the weight matrix of the $l$-th layer, and $\sigma$ is the nonlinear activation function, such as ReLU. When $l = 0$, $H^{(0)} = X^{(k)}$. Note that when illustrating the GCN layers, we omit the index of the omics data for simplicity, without causing ambiguity. In GCN, we use the symmetrically normalized adjacency matrix [16] $\tilde{A}$ where $D$ is degree matrix, $D_{ii} = \sum_j A_{ij}$ and $I$ is identity matrix, which add self-loop to include the node's own features, as shown in Eq. 5.

$$\tilde{A} = D^{-\frac{1}{2}}(A + I)D^{-\frac{1}{2}} \quad (5)$$

After the GCN, two multi-layer perceptron (MLP) layers are employed to convert the extracted features by GCN to the probability of AD. Formally, the MLPs for $k$-th omics is denoted as $f^{(k)}$. The predicted probability of AD diagnosis is then denoted as shown in Eq. 6.

$$\hat{y}_i = f^{(k)}(H^{(L)}) \quad (6)$$

where $\hat{y}_i \in \mathbb{R}^2$, representing the probability of normal control and AD.

**2.2 Models for multi-omics integration**

Selecting the optimal combination of omics modalities for a bi-view model is crucial to enhance classification performance in multi-omics data analysis. Existing methods for biomedical classification tasks utilizing multi-view data often either directly concatenate features from different views or learn to fuse data by assigning weights to each view, combining features in a low-level feature space. However, properly aligning various views without introducing negative effects remains a significant challenge.

To address this issue, the View Correlation Discovery Network (VCDN) [17] is employed in SGUQ to exploit higher-level cross-omics correlations in the label space. Different types of omics data provide unique class-level distinctions, and VCDN is designed to learn both intra-view and cross-view correlations at this higher level.

In the context of determining the best modality combination for the bi-view model, we utilize VCDN to integrate different pairs of omics data types (mRNA expression + DNA methylation, mRNA expression + miRNA expression, or DNA methylation + miRNA expression). For each sample, we obtain the predicted label distributions from two different omics modalities, denoted as $\hat{y}_i^{(1)}$ and $\hat{y}_i^{(2)}$ for the $i$-th sample using $f^{(1)}$ and $f^{(2)}$ with corresponding GCNs. We then construct a cross-omics discovery tensor $C_i \in \mathbb{R}^{c \times c}$, where each entry is calculated as shown in Eq. 7.

$$C_{i,a_1,a_2} = \hat{y}_{i,a_1}^{(1)} \hat{y}_{i,a_2}^{(2)} \tag{7}$$

Here, $\hat{y}_{i,a}^{(k)}$ denotes the $a$-th entry of the predicted label distribution from the $k$-th modality, and $c$ represents the number of classes. This tensor captures the interactions between the predictions of the two modalities, effectively revealing latent cross-view label correlations.

The tensor $C_i$ is then reshaped into a $c^2$-dimensional vector $c_i$ and fed into the VCDN for the final prediction. VCDN is implemented as a fully connected network with an output dimension equal to the number of classes $c$. The loss function for VCDN is defined as in Eq. 8.

$$L_{VCDN} = \sum_{i=1}^{n} L_{CE}(VCDN(c_i), y_i) \tag{8}$$

where $L_{CE}$ represents the cross-entropy loss function, $y_i$ is the ground truth of the $i$-th sample.

By utilizing VCDN to integrate initial predictions from different pairs of omics data, we can assess which modality combinations yield the best classification performance in the bi-view model. The final prediction is based on both the omics-specific predictions and the learned cross-omics label correlation knowledge. The loss function can be written in Eq. 9.

$$L = L_{GCN}^1 + L_{GCN}^2 + L_{VCDN} \tag{9}$$

where $L_{GCN}^{(k)}$ is the cross-entropy loss function measured by the GCN predicted label and ground truth. While the bi-view model aims to improve prediction confidence by integrating two modalities, there may still be instances where the model's predictions are not sufficiently confident. In such cases, we enhance the model by incorporating a third modality, effectively transitioning to a tri-view model. The tri-view model is constructed and trained in the same manner, as shown in Eq. 10.

$$C_{i,a_1,a_2,a_3} = \hat{y}_{i,a_1}^{(1)} \hat{y}_{i,a_2}^{(2)} \hat{y}_{i,a_3}^{(3)} \tag{10}$$

For the three modalities, we construct a tensor $C_j \in \mathbb{R}^{c \times c \times c}$. This tensor captures the complex interactions among the three modalities at the class level. It was then reshaped into a $c^3$-dimensional vector and fed into an extended VCDN tailored to handle the additional modality. And the total loss function is defined as shown in Eq. 11.

$$L = \sum_{k=1}^{3} L_{GCN}^{(k)} + L_{VCDN} \tag{11}$$

**Model training strategies.** During the training process, we start by pretraining each omics-specific GCN individually to achieve a good initialization for the networks. Then, in each epoch of the training phase, we perform an alternating optimization:

1) Update Omics-Specific GCNs: We first fix the parameters of the VCDN and update each omics-specific $GCN_i$ for $i = 1,2,3$, minimizing the loss function $L_{GCN}$. This step adjusts the omics-specific networks to better capture the features of their respective data types.

2) Update VCDN: Next, we fix the parameters of the omics-specific GCNs and update the VCDN to minimize the same loss function $L$ in Eq. 9 for bi-view model and Eq. 11 for tri view model. This step allows the VCDN to learn the cross-omics label correlations based on the updated outputs from the GCNs.

3) We repeat this alternating optimization of the omics-specific GCNs and the VCDN until the model converges to a stable solution.

**2.3 Uncertainty quantification for multi-omics data integration**

In our multi-omics data analysis framework, we incorporate uncertainty quantification to evaluate the confidence of the model's predictions. High uncertainty indicates lower confidence, suggesting that the model's predictions for certain samples may be less reliable. Incorporating uncertainty quantification allows us to identify reliable predictions: Predictions with low uncertainty are more likely to be accurate, enhancing the trustworthiness of the model in disease diagnosis.

In detail, we quantify uncertainty by assessing the consistency of the model's predictions over multiple trials and calculating the standard deviation of the predicted probabilities. Our model performs multiple independent predictions with each sample. Specifically, we conduct $T$ trials, where in each trial, the model experiences variations due to random initialization or stochastic training processes. The final predicted label for each sample is determined by a majority voting mechanism across these trials.

To quantify uncertainty, we calculate the standard deviation of the predicted probabilities from the multiple trials for each sample. Let $p_i^{(t)} = argmax\ \hat{y}_i$ represent the predicted probability for patient $i$ in trial $t$, where $t = 1,2,\ldots,T$. The mean predicted AD probability for patient $i$ is denoted in Eq. 12.

$$\bar{p}_i = \frac{1}{T}\sum_{t=1}^{T} p_i^{(t)} \qquad (12)$$

The standard deviation $\sigma_i$ of the predicted probabilities is calculated as shown in Eq. 13, which is denoted as the uncertainty of the model prediction.

$$\sigma_i = \sqrt{\frac{1}{T-1}\sum_{t=1}^{T}(p_i^{(t)} - \bar{p}_i)^2} \qquad (13)$$

A higher $\sigma_i$ indicates greater variability in the model's predictions for patient $i$, reflecting higher uncertainty. By quantifying uncertainty through ensemble predictions and standard deviation calculations, we can distinguish between confident and uncertain predictions, thereby improving the overall reliability of the model. High uncertainty in specific patients may indicate the need for additional data collection or alternative experimental approaches to reduce uncertainty. Providing uncertainty estimates alongside predictions aids clinicians in assessing the trustworthiness of the model's recommendations, facilitating more informed and safer clinical decisions.

Determining the best modality combination involves evaluating various pairs of omics data to identify which ones complement each other most effectively. By starting with the best-performing single modality and progressively adding other modalities, we can assess the impact on prediction confidence and accuracy. If the bi-view model's predictions remain uncertain, incorporating a third modality can provide additional information to resolve ambiguities.

## 2.4 Determine the optimal uncertainty threshold

The staged method is a progressive multi-view learning approach designed to improve classification performance by incorporating additional data modalities when predictions lack confidence. This methodology is particularly useful in AD diagnosis, where different biological data types provide complementary information. The method operates by introducing new data views in stages based on the uncertainty of predictions at each stage.

The staged approach aims to mitigate uncertainty by sequentially incorporating additional data types when the confidence of predictions falls below a predefined threshold. Initially, a model is trained using a single data view (e.g., mRNA data). If the uncertainty in the predictions exceeds a threshold, additional views, such as DNA methylation or miRNA data, are introduced to improve the model's confidence. The process continues until the uncertainty is sufficiently reduced or all available views have been utilized.

Key steps of our SGUQ includes:

1) Initial View (Single-View Classification): The proposed SGUQ begins by training and evaluating a machine learning model using a single type of biological data, such as mRNA. For each patient or sample, the model generates predictions along with an uncertainty measure (e.g., standard deviation of predictions). If the uncertainty for a given prediction is lower than a predefined threshold ($t_1$), the result is considered confident, and no further data views are needed. If the uncertainty exceeds the threshold, the sample is flagged as unconfident, prompting the model to incorporate an additional view in the next stage. The model employed in the single-view classification is defined in Section 2.1.

2) Second View (Bi-Modal Classification): For samples where the initial single-view prediction is uncertain, a second type of data, such as DNA methylation, is added to the model. This bi-modal approach combines the initial mRNA data with the DNA methylation data, providing a more comprehensive analysis. A new threshold ($t_2$) is applied to assess the uncertainty of the bi-modal predictions. If the uncertainty for a sample drops below $t_2$, the prediction is accepted as confident at this stage. If uncertainty remains high, the sample moves to the third stage for further refinement.

3) Third View (Tri-Modal Classification): For cases where neither the single-view nor the bi-modal predictions meet the confidence criteria, a third data type, such as miRNA, is added to the model. This tri-modal classification uses all available data to make a final prediction. By integrating multiple data sources, the model can significantly reduce uncertainty and improve overall accuracy.

At this stage, all patients are classified with the highest level of confidence possible based on the available data.

**Threshold Optimization:** The staged method relies on the careful selection of thresholds ($t_1$ and $t_2$) to determine when additional data views should be introduced. These thresholds are optimized through an iterative process, where different threshold values are tested, and the accuracy of the predictions is evaluated. The thresholds that yield the best performance, as measured by overall accuracy and other metrics such as F1-score or ROC-AUC, are selected. Formally, the uncertainty threshold selection algorithm is shown in Algorithm 1.

**Algorithm 1**. Threshold Selection for Staged Classification.

---

**Input**: single-, bi- and tri-view uncertainties, threshold_range_single, threshold_range_bi
**Output**: Optimal thresholds $t_1$ and $t_2$ for classification
1: Initialize *best_accuracy* ← 0, *best_t1* ← 0, *best_t2* ← 0
2: Define threshold range uni and threshold range bi as ranges for $t_1$ and $t_2$
3: **for** each $t_1$ in *threshold_range_uni* **do**
4:     **for** each $t_2$ in *threshold_range_bi* **do**
5:         Initialize *classified_patients* ← [ ]
6:         Initialize *unconfident_patients_stage1* ← [ ]
7:         Initialize *unconfident_patients_stage2* ← [ ]
        ▷ Stage 1: Single-View Classification
8:         **for** each *patient* **do**
9:             **if** *patient*.std ≤ $t_1$ **then**
10:                 *classified_patients*.append(patient)
11:             **else**:
12:                 *unconfident_patients_stage1*.append(patient)
13:             **end if**
14:         **end for**
        ▷ Stage 2: Bi-View Classification
15:         **for** each *patient* where *patient* in *unconfident_patients_stage1* **do**
16:             **if** *patient*.std ≤ $t_2$ **then**
17:                 *classified_patients*.append(patient)
18:             **else**
19:                 *unconfident_patients_stage2*.append(patient)
20:             **end if**
21:         **end for**
        ▷ Stage 3: Tri-View Classification
22:         **for** each *patient* where patient in *unconfident_patients_stage2* **do**
23:             *classified_patients*.append(*patient*)
24:         **end for**
        ▷ Evaluate accuracy for current thresholds
25:         *total_accuracy* ← calculate_accuracy(*classified_patients*)
26:         **if** *total_accuracy* > *best_accuracy* **then**
27:             *best_accuracy, best_t1, best_t2* ← *total_accuracy*, $t_1$, $t_2$
28:         **end if**
29:     **end for**
30: **end for**
31: **return** *best_t1, best_t2*

---

**Explanation of the Algorithm**: In Algorithm 1, the input contains single-, bi- and tri-view uncertainties, representing the predicted results for single-, bi- and tri-view models along with the uncertainty quantifications results. The threshold_range_single and threshold_range_bi start from the minimum uncertainty and end at the maximum uncertainty, divided into 100 steps. Then, the algorithm iterates over each possible combination of thresholds $t_1$ and $t_2$ (Lines 3 and 4). At Stage 1 (Lines 8 to 14) - Single-View Classification: For each threshold $t_1$, patients are classified using the single-view model (e.g., mRNA data). If the uncertainty for a patient is below $t_1$, the prediction is considered confident, a definitive result is obtained, and no further tests are required. If not, the patient is flagged for Stage 2. At Stage 2 (Lines 15 to 21) - Bi-View Classification: For patients with unconfident predictions from Stage 1, the model introduces a second data view (e.g., DNA methylation). If the uncertainty for these patients is below $t_2$, the prediction is considered confident. If not, they proceed to Stage 3. At Stage 3 (Lines 22 to 24)- Tri-View Classification: Unconfident patients from Stage 2 are classified using all available data views (e.g., mRNA + DNA methylation + miRNA). After all patients reach a definitive result, the total accuracy is calculated. If the new accuracy exceeds the best observed accuracy, the algorithm updates the best accuracy and stores the corresponding thresholds $t_1$ and $t_2$.

### 2.5 Evaluation metrics

To compare the performance of the proposed SGUQ method with other approaches for binary classification of Alzheimer's disease (AD), where patients are categorized into AD and NC groups, we employ accuracy (ACC), F1-score (F1), and the area under the receiver operating characteristic curve (AUC) as evaluation metrics.

## 3. Experimental results

### 3.1 Dataset and enrolled subjects

The ROSMAP dataset is a comprehensive resource derived from two longitudinal clinical studies: the Religious Orders Study (ROS) and the Memory and Aging Project (MAP) [18]. This dataset, which contains 351 subjects, is used for distinguishing AD subjects from normal controls [19,20]. These studies aim to understand the clinical, pathological, and molecular underpinnings of Alzheimer's disease and other neurodegenerative disorders. Participants in these studies are older adults who undergo annual clinical evaluations and agree to organ donation at the time of death, allowing for extensive post-mortem analyses. The main characteristics of the ROSMAP dataset include high-throughput molecular data (i.e. multi-omics data) generated from brain tissue samples, enabling integrative analyses.

In our study, we utilize three types of omics data from the ROSMAP dataset to train our models, capturing different layers of biological information related to Alzheimer's disease:

1) mRNA Expression Data (Transcriptomic): mRNA expression data quantifies the levels of messenger RNA transcripts in cells, reflecting gene activity and functional output. Alterations in gene expression patterns are associated with disease states. In Alzheimer's disease, specific genes may be upregulated or downregulated, contributing to disease progression [21]. mRNA expression data help identify these genes and understand their roles.
2) DNA Methylation Data (Epigenomics): DNA methylation involves the addition of methyl groups to DNA molecules, typically at cytosine-phosphate-guanine (CpG) sites, affecting gene expression without changing the DNA sequence. Epigenetic modifications like DNA methylation play crucial roles in regulating gene expression and can be influenced by environmental factors and aging. Abnormal methylation patterns have been linked to Alzheimer's disease, making this data valuable for uncovering epigenetic contributions to the disease [22].
3) miRNA Expression Data (Non-Coding RNA Transcriptomics): MicroRNAs (miRNAs) are small non-coding RNA molecules that regulate gene expression post-transcriptionally by binding to target mRNAs, leading to their degradation or inhibition of translation. miRNAs are key regulators of gene expression and are involved in various cellular processes, including neuronal development and function. Dysregulation of miRNAs has been implicated in Alzheimer's disease, influencing pathways related to neurodegeneration [23].

We employed the pre-processed dataset from MOGONET [12] to conduct the experiments in this paper. This public dataset includes 169 normal control patients and 182 AD patients. The included number of omics features are shown in Table 1.

**Table 1.** The ROSMAP dataset is for the classification of Alzheimer's disease (AD) patients and normal control (NC).

| Dataset | Categories | Number of Features for training mRNA, DNA meth, miRNA |
|---|---|---|
| ROSMAP | 2 (NC: 169 patients, AD: 182 patients) | 200, 200, 200 |

### 3.2 Model implementation and omics selection

We implemented our model using Python 3.8 and Pytorch1.10. The number of trials, i.e. $T$ in Eq. 12 was set as 10. The average number of edges retained per node (including self-connections) defined in Eq. 2 was set as $K = 2$ for each experiment. In addition, the number of GCN layers, i.e. $L$, was set as 3. Each GCN model contains 3 GCN layers, with 200, 200 and 100 hidden units, respectively.

We evaluated the model's performance using ACC, AUC, and F1-score, as described in Section 2.5. We first iteratively performed AD classification using the single-view, bi-view and tri-view classification models using different combinations of the omics data, as shown in Table 2.

**Table 2.** Performance and the average uncertainty for AD classification using the proposed SGUQ. The bold texts indicate the achieved best performance under corresponding scenarios.

| Model | Enrolled omics data | ACC ↑ | F1-score ↑ | AUC ↑ | Average uncertainty ↓ |
|---|---|---|---|---|---|
| Single-view (Stage 1) | mRNA | **0.8018** | **0.8073** | **0.8019** | **0.0781** |
| | DNAmeth | 0.6320 | 0.6213 | 0.6340 | 0.0790 |
| | miRNA | 0.6981 | 0.7142 | 0.6969 | 0.0991 |
| Bi-view (Stage 2) | mRNA+DNAmeth | **0.8396** | **0.8349** | **0.8418** | 0.0979 |
| | mRNA+miRNA | 0.8301 | 0.8301 | 0.8313 | 0.1417 |
| | DNAmethy+miRNA | 0.7169 | 0.7000 | 0.7201 | **0.0903** |
| Tri-view (Stage 3) | mRNA+DNAmeth+miRNA | **0.8584** | **0.8571** | **0.8600** | 0.1588 |

According to Table 2, the single-view models rely on individual omics data types—mRNA, DNA methylation (DNAmeth), and miRNA—for AD classification. Among these, the mRNA-only model achieves the best performance, with an accuracy of 0.8018, an F1-score of 0.8073, and an AUC of 0.8019. This shows that mRNA data alone provides relatively strong predictive capability for distinguishing AD patients from normal controls. In contrast, models based solely on DNA methylation and miRNA perform worse, with the DNA methylation model achieving the lowest overall performance (ACC = 0.6320, F1 = 0.6213, AUC = 0.6340). The miRNA-only model performs better than DNA methylation but remains less effective than mRNA, with an accuracy of 0.6981 and an F1-score of 0.7142. These results suggest that while mRNA data is more informative for AD classification at Stage 1, DNA methylation and miRNA may contribute important, complementary information that cannot be captured in isolation.

The bi-view models, which combine two types of omics data, significantly improve performance compared to single-view models. The mRNA+DNAmeth model achieves the best bi-view performance, with an accuracy of 0.8396, an F1-score of 0.8349, and an AUC of 0.8418. These results demonstrate that integrating DNA methylation data with mRNA boosts the model's ability to classify AD patients more accurately, likely due to the complementary nature of these data types. The reduction in average uncertainty to 0.0979 also shows increased confidence in the predictions compared to mRNA+miRNA model. This indicates that while miRNA data improves predictive accuracy, it introduces greater uncertainty than DNA methylation when combined with mRNA. Finally, the combination of DNA methylation and miRNA provides moderate performance (ACC = 0.7169), indicating that while useful, this pairing does not fully capture the biological complexity of AD.

The tri-view model, which integrates all three omics data types (mRNA, DNA methylation, and miRNA), achieves the highest overall performance, with an accuracy of 0.8584, an F1-score of 0.8571, and an AUC of 0.8600. The combination of all available data types allows the model to capture a more comprehensive biological signal, significantly improving classification accuracy and robustness.

The choice of omics data for AD classification depends on the desired trade-off between accuracy and complexity. For simpler models with reasonable accuracy and low uncertainty, using mRNA data alone may suffice. However, for improved performance, especially in scenarios where the classification task is more challenging, combining mRNA with DNA methylation or miRNA is recommended. The tri-view model, while yielding the highest accuracy, may introduce higher uncertainty due to the complexity of integrating multiple data types. Therefore, this model is best suited for cases where maximizing accuracy is crucial, and higher computational complexity can be tolerated.

**3.3 Model performance and comparison**

Additionally, we compared the proposed SGUQ with existing methods for AD classification using the ROSMAP dataset with the same train/test subject separation. For the performance evaluations using ROSMAP multi-omics datasets, we compared the proposed method with the following 20 existing classification algorithms. The performance comparison is shown in Table 3.

- Five machine learning algorithms. K-nearest neighbor classifier (**KNN**) [24], Support Vector Machine (**SVM**) [25], Linear Regression with L1 regularization (**LR**), Random Forest (**RF**) [26] and fully connected neural networks (**NN**).
- Adaptive group-regularized ridge regression (**GRridge**) [9]. GRridge is a method for adaptive group-regularized ridge regression with group-specific penalties for high-dimensional data classification.
- Block partial least squares discriminant analysis (**BPLSDA**) and Block sparse partial least squares discriminant analysis (BSPLSDA) [10]. BPLSDA extends sparse generalized canonical correlation analysis to a classification framework, and BSPLSDA adds sparse constraints to BPLSDA.
- Multi-Omics Graph cOnvolutional NETworks (**MOGONET**) [12]: MOGONET jointly explores omics-specific learning using graph convolution network and cross-omics correlation learning using view correlation discovery network for multi-omics integration and classification.
- Trusted multi-view classification (**TMC**) [27]. TMC dynamically computes the trustworthiness of each modality for different subjects with reliable integration for multi-view classification.
- Concatenation of final multimodal representations (**CF**) [28]. CF performs multi-view information fusion based on late fusion and compactness-based fusion.
- Gated multimodal units for information fusion (**GMU**) [29]. GMU employs the gates for selecting the most important parts of the input of each modality to correctly generate the desired output.
- Multi-modality dynamic fusion (**MMDynamic**) [30]. MMDynamic employs the feature-level and modality-specific gates for multi-modality data fusion.
- Multi-Level Confidence Learning for Trustworthy Multimodal Classification (**MLCLNet**) [31]: MLCLNet employs a feature confidence learning mechanism to suppress redundant features and a graph convolution network to learn the corresponding structure for multi-view information fusion.
- Multiview canonical correlation analysis (**MCCA**) [32]. MCCA extends the canonical correlation analysis (CCA) into multi-view settings. CCA is a typical subspace learning algorithm, aiming at finding the pairs of projections from different views with the maximum correlations. For more than 2 views, MCCA optimizes the sum of pairwise correlations. For samples with missing modalities, the trained MCCA projects the existing modalities into the subspace as the feature representations.
- Kernel CCA (**KCCA**) [33]. KCCA extends MCCA by adding kernel techniques.

- Kernel generalized CCA (**KGCCA**) [34]. KGCCA extends KCCA with a prior-defined graph between different modalities.
- Sparse CCA (**SCCA**) [35]. SCCA extends CCA with modality-specific sparse penalty.
- Multi-view Variation AutoEncoder (**MVAE**) [36,37]. MVAE extends variational autoencoders for latent feature extraction and employs the product of the multivariate Gaussian distributions for information fusion. For samples with missing modalities, the trained MVAE generates latent representation by the product of the latent features from the existing modalities.
- Cross Partial Multi-View Networks (**CPM**) [38]. CPM performs feature embedding using in-complete multi-view data by focusing on the completeness and versatility of the feature embedding.
- Dual Contrastive Prediction (**DCP**) [39]. DCP employs maximizes the conditional entropy through dual prediction to recover the missing views and employs dual contrastive loss to learn the consistent representation among different modalities.

**Table 3**. Comparison with the state-of-the-art algorithms for multi-omics data classification. The bold texts indicate the best performance.

| Methods | ACC ↑ | F1-score ↑ | AUC ↑ |
| --- | --- | --- | --- |
| KNN | 0.657±0.036 | 0.671±0.045 | 0.709±0.045 |
| SVM | 0.770±0.024 | 0.778±0.026 | 0.770±0.026 |
| LR | 0.694±0.037 | 0.730±0.035 | 0.770±0.035 |
| RF | 0.726±0.029 | 0.734±0.019 | 0.811±0.019 |
| NN | 0.755±0.021 | 0.764±0.025 | 0.827±0.025 |
| GRridge | 0.760±0.034 | 0.769±0.023 | 0.841±0.023 |
| BPLSDA | 0.742±0.024 | 0.755±0.025 | 0.830±0.025 |
| BSPLSDA | 0.753±0.033 | 0.764±0.021 | 0.838±0.021 |
| MOGONET | 0.815±0.023 | 0.821±0.012 | 0.874±0.012 |
| TMC | 0.825±0.009 | 0.823±0.006 | 0.885±0.006 |
| CF | 0.784±0.011 | 0.788±0.005 | 0.880±0.005 |
| GMU | 0.776±0.025 | 0.784±0.016 | 0.869±0.016 |
| MMDynamics | 0.842±0.013 | 0.846±0.007 | **0.912±0.007** |
| MLCLNet | 0.844±0.015 | 0.852±0.015 | 0.893±0.011 |
| KGCCA | 0.695±0.000 | 0.680±0.000 | 0.697±0.000 |
| SCCA | 0.810±0.000 | 0.811±0.000 | 0.810±0.000 |
| MVAE | 0.752±0.026 | 0.750±0.031 | 0.753±0.025 |
| CPM | 0.742±0.024 | 0.729±0.027 | 0.741±0.024 |
| DCP | 0.785±0.000 | 0.805±0.000 | 0.786±0.000 |
| LHGN | 0.759±0.015 | 0.756±0.023 | 0.760±0.015 |
| Proposed | **0.858±0.006** | **0.857±0.009** | 0.860±0.000 |

The results presented in Table 3 demonstrate a comparative analysis of various state-of-the-art algorithms for multi-omics data classification. Among the classical machine learning models, KNN exhibited the lowest performance, with an ACC of 0.657 and an AUC of 0.709, indicating its limitations in handling complex multi-omics data. SVM and Logistic Regression performed moderately, with ACCs of 0.770 and 0.694, respectively, while Random Forest and Neural Networks improved performance, achieving ACCs of 0.726 and 0.755, respectively.

In contrast, the proposed method outperformed all existing algorithms, achieving an ACC of 0.858, an F1-score of 0.857, and an AUC of 0.860. Notably, MMDynamics and MLCLNet also performed well, with ACCs of 0.842 and 0.844, respectively. However, the proposed method's consistent superiority across all metrics underscores its effectiveness in classifying multi-omics data. This analysis demonstrates the importance of utilizing advanced approaches in the classification of complex biological data, emphasizing the proposed method's significant contribution to the field.

**3.4 Analysis of the SGUQ for cost-effective decision-making**

In the process of selecting optimal thresholds for the ROSMAP dataset, we focused on two critical thresholds: $t_1$ and $t_2$. The first threshold, $t_1$, is specifically associated with single-view data, while the second threshold, $t_2$, pertains to bi-view data. The selected values for these thresholds were determined through extensive evaluation of classification performance. For the ROSMAP dataset, the optimal thresholds were identified as follows:

- **Best Threshold** $t_1$: 0.0527
- **Best Threshold** $t_2$: 0.1024

These thresholds represent the points at which the classification model achieved the highest accuracy in distinguishing between classes. By applying these thresholds, the model effectively classified instances into confident predictions, thereby enhancing the overall performance of multi-omics data classification in the context of AD diagnosis and lowering the clinical costs.

Through analysis of the dataset, we found that 46.23% of the samples can be reliably predicted using only single-modal omics data (mRNA), while an additional 16.04% of the samples can achieve reliable predictions when combining two omics data types (mRNA + DNA methylation). This finding triggers our further discussion.

Omics data collection is typically expensive, especially for multi-omics data. If nearly half of the samples can be accurately predicted using only mRNA data, it implies significant reductions in data acquisition and experimental costs. At the same time, focusing limited resources on samples that require additional omics data improves resource utilization efficiency. Initially, we used low-cost mRNA data for preliminary prediction. For samples with high uncertainty, collect DNA methylation data for further analysis. This staged strategy effectively reduces overall costs. Concentrating efforts and resources on samples with high prediction uncertainty also helps improve the overall performance and reliability of the model.

For a large number of samples, reliable predictions can be obtained using a single omics dataset, speeding up the diagnostic and therapeutic decision-making process. Reducing the number and types of tests required for patients lowers the medical burden and waiting time. Meanwhile, decisions about whether additional omics data are needed can be made based on the prediction uncertainty of each sample, enabling customized data collection and truly personalized medicine. For patients requiring multi-omics data, more in-depth analysis and precise treatment recommendations can be provided, optimizing their treatment plans.

Our results demonstrate the predictive power of mRNA data, which is sufficient to provide accurate predictions for a large number of samples, emphasizing its importance in biomarker research. For samples requiring additional omics data, it demonstrates the value of multi-omics integration in improving prediction accuracy. In-depth research on samples that require multi-omics data for accurate prediction may reveal new biological pathways and disease mechanisms. These findings can help researchers identify the most promising areas for research and optimize resource allocation.

We further visualized the distribution of the uncertainty quantification results for each model, as shown in Figure 2.

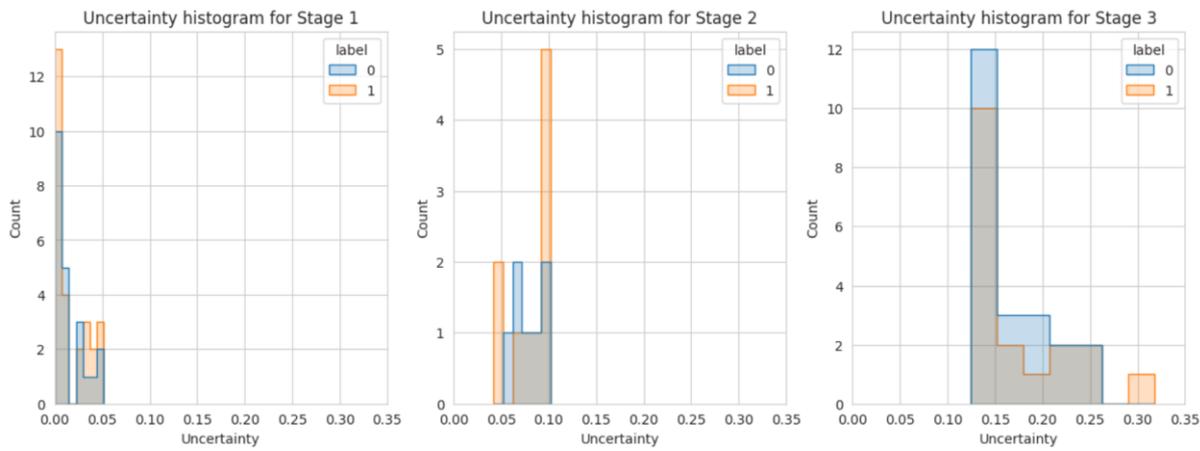

**Figure 2.** Distributions of the uncertainty quantification results for single-, bi- and tri-view AD classification model. The label 0 indicates the normal control while label 1 indicates the AD patients.

Figure 2 illustrates the uncertainty histograms across the three stages of the proposed SGUQ for AD diagnosis. In **Stage 1**, where only mRNA data is used, most predictions have very low uncertainty, indicating that a significant portion of samples can be reliably classified with just single-modal data. As the model progresses to **Stage 2**, incorporating DNA methylation data, the uncertainty slightly increases, but the predictions remain relatively confident for additional samples that require this second data type. In **Stage 3**, where miRNA data is introduced, the uncertainty increases further, particularly for samples that are more challenging to classify, even with all three omics data types. This staged approach demonstrates the model's ability to balance diagnostic accuracy with cost-effectiveness by progressively incorporating more omics data only when necessary.

## 4. Discussion

### 4.1 Benefits of Adding Uncertainty Module

In multi-omics data analysis, incorporating an uncertainty module significantly enhances the model's performance and application value. By quantifying and analyzing the uncertainty of model predictions, we gain deeper insights into the model's behavior, optimize resource allocation, enhance robustness, and provide guidance for future research.

By setting appropriate uncertainty thresholds, the model can automatically distinguish between high-confidence and low-confidence predictions, enhancing prediction credibility. High-confidence predictions imply that the model has strong confidence in its classification of these samples, which allows us to place greater trust in these results. In critical application areas such as disease diagnosis and personalized medicine, incorrect predictions can have severe consequences. By focusing on high-confidence predictions, we can reduce the risk of misclassification and improve the model's reliability in practical applications. When the model only outputs high-confidence predictions, the user's (e.g., clinicians or researchers) trust in the model is also increased. This contributes to the acceptance and effectiveness of the model.

For further research, samples with high uncertainty may reflect complex, biologically important patterns in the data that have not yet been fully understood. These samples may contain unknown disease subtypes, new biomarkers, or undiscovered biological mechanisms. Understanding which samples have high uncertainty can guide future data collection and experimental design. For example, additional sample collection, obtaining more omics-level data, or using different experimental techniques can be considered for these samples. By identifying and thoroughly analyzing these high-uncertainty samples, researchers can discover new scientific questions and formulate new hypotheses, thereby driving advancements in the field.

### 4.2 Staged model in clinical decision making

Data acquisition and processing costs are often very high in multi-omics analysis. By using uncertainty thresholds, we can adopt a staged analysis strategy, optimizing resource utilization. For stage one, we use a single omics dataset for preliminary predictions, which is fast and cost-effective. For stage two, samples with high uncertainty incorporate more omics data for deeper analysis to improve the prediction accuracy of these samples. This strategy allows us to reduce the reliance on expensive data without significantly compromising the overall model performance, thereby lowering the overall research cost.

The overall accuracy of the model's predictions is then evaluated for each combination of thresholds from different stages. The combination that yields the highest accuracy is selected as the optimal set of thresholds. This ensures that the model achieves the best possible performance by effectively balancing confident and uncertain predictions.

## 5. Conclusion

In this study, we introduced a novel staged deep learning approach that leverages multi-omics data for improved AD diagnosis. By utilizing high-throughput omics technologies—specifically genomics, transcriptomics, and epigenomics—we addressed the limitations of conventional AI methodologies that necessitate the completion of all omics data before diagnosis. Our innovative approach incorporates mRNA data as the initial stage, progressively integrating DNA methylation and miRNA data only when necessary. This staged method not only reduces clinical costs and patient exposure to potentially harmful tests but also enhances diagnostic accuracy. The experimental results demonstrated the effectiveness of our approach, achieving an accuracy of 0.858 on the ROSMAP dataset and significantly outperforming existing classification methods. Notably, our model was able to reliably predict outcomes for 46.23% of samples using single-modal omics data (mRNA) alone, with an additional 16.04% of samples benefiting from the combination of mRNA and DNA methylation data. These findings highlight the potential of our staged deep learning framework not only for AD diagnosis but also for broader applications in clinical decision-making using multi-view data. Our work paves the way for more efficient and cost-effective strategies in the detection and management of Alzheimer's Disease and underscores the importance of adopting flexible methodologies in the era of personalized medicine.

**Declaration of Competing Interest**

The authors declare that they have no known competing financial interests or personal relationships that could have appeared to influence the work reported in this paper.

**Credit authorship contribution statement**

Liang Tao: Conceptualization, methodology, manuscript writing.

Yixin Xie: Methodology, manuscript review

Hong-Wen Deng: Conceptualization, methodology, manuscript writing, and review.

Weihua Zhou: Conceptualization, manuscript writing, and review.

Chen Zhao: Supervision, project administration, software, funding acquisition, manuscript writing, and review.

**Acknowledgement**

This research was supported by a research seed fund from Kennesaw State University and three NIH grants (1R15HL172198, 1R15HL173852, and U19AG055373).